\documentclass[aps,prl,twocolumn,groupedaddress,showpacs]{revtex4}

\usepackage{graphicx}
\usepackage{bm}

\newcommand{\kBEDT}[1]{$\kappa$-(BEDT-TTF)$_2$Cu[N(CN)$_2$]#1}
\newcommand{\kX}{$\kappa$-(ET)$_2X$}
\newcommand{\kCl}{$\kappa$-(ET)$_2$Cl}
\newcommand{\kBr}{$\kappa$-(ET)$_2$Br}
\newcommand{\muB}{\ensuremath{\mu_{\rm B}}}

\begin{document}

\title{Spin diffusion and magnetic eigenoscillations confined to single
molecular layers in the organic conductors \kBEDT{$X$}\ ($X$=Cl, Br)}

\author{\'Agnes Antal}
\author{Titusz Feh\'er}
\author{Andr\'as J\'anossy}\email[]{atj@szfki.hu}
\author{Erzs\'ebet T\'atrai-Szekeres}
\author{Ferenc F\"ul\"op}
\affiliation{Budapest University of Technology and Economics, Department
of Experimental Physics and Condensed Matter Physics Research Group of
the Hungarian Academy of Sciences, H-1521 Budapest, Hungary}

\date{\today}

\begin{abstract}
The layered organic compounds, \kBEDT{$X$} ($X$=Cl, Br) are metals at
ambient temperatures. At low temperatures, the Cl compound is a weakly
ferromagnetic Mott insulator while the isostructural Br compound is a
superconductor. We find by conduction electron spin resonance (CESR) and
antiferromagnetic resonance (AFMR) an extreme anisotropy of spin
transport and magnetic interactions in these materials. In the metallic
state spin diffusion is confined to single molecular layers within the
spin lifetime of $10^{-9}\rm\,s$. Electrons diffuse several hundreds of
nm without hopping to the next molecular layer. In the magnetically
ordered insulating phase of the Cl compound we observe and calculate the
four AFMR modes of the weakly coupled single molecular layers. The
inter-plane exchange field is comparable or less than the typically
$1\rm\,mT$ dipolar field and almost $10^6$ times less than the
intra-layer exchange field.
\end{abstract}

\pacs{71.20.-b,72.25.-b,75.30.-m,76.30.Pk}

\keywords{antiferromagnetic resonance, AFMR, conduction electron spin
resonance, CESR, organic salt, ET, BEDT, BEDT-TTF, layered, quasi
two-dimensional, anisotropic, spin lifetime, spin transport,
spintronics, Dzyaloshinskii-Moriya, Dzialoshinskii-Moriya,
Dzyaloshinsky-Moriya}

\maketitle


In quasi two-dimensional conductors the anisotropy may be so high that
the $1/\tau_\parallel$ intra-layer momentum scattering rate exceeds the
tunneling rate from one atomic or molecular layer to the next. In such
conductors charge transport perpendicular to the layers is incoherent
\cite{Kartsovnik2004}. The condition for conduction electron spin
transport to be two dimensional is more stringent: spins in adjacent
molecular layers must not mix within the spin lifetime, $T_1$. In
organic layered conductors $T_1$ is typically $1\rm\,ns$, several orders
of magnitude longer than $\tau_\parallel$, and the anisotropy of the
momentum scattering has to be extreme for a two-dimensional spin
diffusion. 

The magnetic resonance experiments described below show that adjacent
layers of the isostructural organic compounds, \kX\ [\kX\ is
\kBEDT{$X$}, $X$=Cl, Br, and
ET=BEDT-TTF=bis(ethylenedithio)tetrathiafulvalene], are to a
large extent magnetically isolated, testifying two-dimensional spin
dynamics. In the metallic state, spin diffusion is confined to single
layers. In the ordered state, magnetic eigenoscillations of adjacent
layers in external fields are almost completely independent, except in
some mode crossing directions. The $10^6$ ratio of intra-plane to
inter-plane exchange fields explains why previous observations
\cite{Ohta1999,Ito2000} were inconclusive. We are not aware of any other
layered crystal in which two-dimensional spin diffusion or independent
oscillations of adjacent magnetic layers was observed.

\begin{figure}
\includegraphics[bb=210 170 730 450,clip,width=\columnwidth]{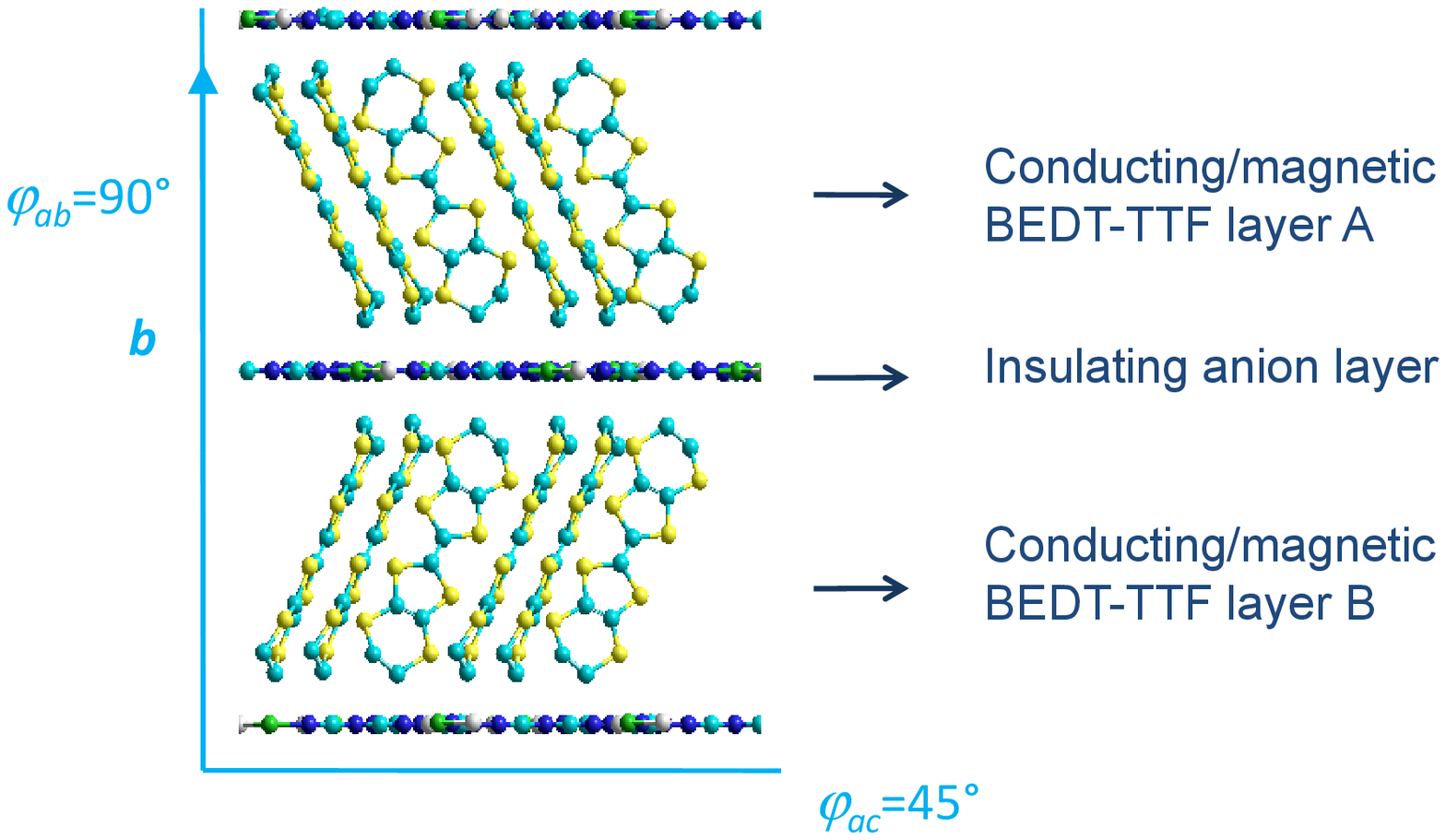}%
\caption{(color online) Crystal structure of the \kX, $X$=Cl, Br layered
compounds. $\varphi_{ab}$ and $\varphi_{ac}$ denote angles from $\bm{a}$
in the $(\bm{a},\bm{b})$ and $(\bm{a},\bm{c})$ planes, respectively.
Electronic overlap between ET molecules in adjacent A and B layers is
small, typically $t_\perp=0.1\rm\,meV$. In the metallic state, the
Larmor frequencies of the two chemically equivalent but magnetically
inequivalent layers, A and B, are different in general magnetic field
orientations.\label{fig:crystal}}
\end{figure}

\kCl\ and \kBr\ have an orthorhombic crystallographic cell
\cite{Kini1990,Williams1990} with two symmetry-related, chemically
equivalent ET layers, A and B, separated by polymeric anion layers
(Fig.~\ref{fig:crystal}). The ET molecules are arranged into dimers with
formally $+e$ charge and 1/2 spin. Inter-dimer overlap is significantly
less than intra-dimer overlap, and electronic bands are half filled. At
high temperatures the conductivity is metallic and very anisotropic:
perpendicular conductivity measured by dc methods is about 1000 times
less than in-plane \cite{Buravov1992}. \kCl\ undergoes a
metal--insulator Mott transition at $27\rm\,K$ and the ground state is a
canted antiferromagnetic insulator \cite{Welp1992}. A small pressure
suppresses the Mott transition in \kCl, which becomes a superconductor
below $12.5\rm\,K$, similarly to \kBr\ at ambient pressure
\cite{Ito1996,Lefebvre2000}.


We first discuss the conduction electron spin resonance (CESR) in the
metallic phase between 45 and $250\rm\,K$. In magnetic fields, $\bm{H}$,
in general directions there are four differently oriented, magnetically
inequivalent ET dimers, A1, A2 and B1, B2. Whether lines of molecules
are resolved in electron spin resonance (ESR) spectra depends on the
electronic overlap. Spin--orbit and crystal field interactions render
the $g$ factor anisotropic and the ESR lines of inequivalent isolated
magnetic molecules are split by $\Delta\omega_{12}=(g_1-g_2)\muB
H/\hbar$, where \muB\ is the Bohr magneton and $g_{1,2}$ are the
effective gyromagnetic factors. The ESR lines of interacting molecules
merge into a single line if their spins are exchanged within a time of
$1/\Delta\omega_{12}$. In quasi one-dimensional and two-dimensional
conductors with inequivalent molecular stacks or layers, motional
averaging into a single ESR line depends on whether or not the
inter-chain or inter-layer tunnelling rate, $\nu_\perp$, exceeds the
difference in Larmor frequencies. A single CESR line appears in most
quasi one-dimensional organic compounds \cite{Tomkiewicz1975}.

\begin{figure}
\includegraphics[width=\columnwidth, trim=0 25 0 55,clip]{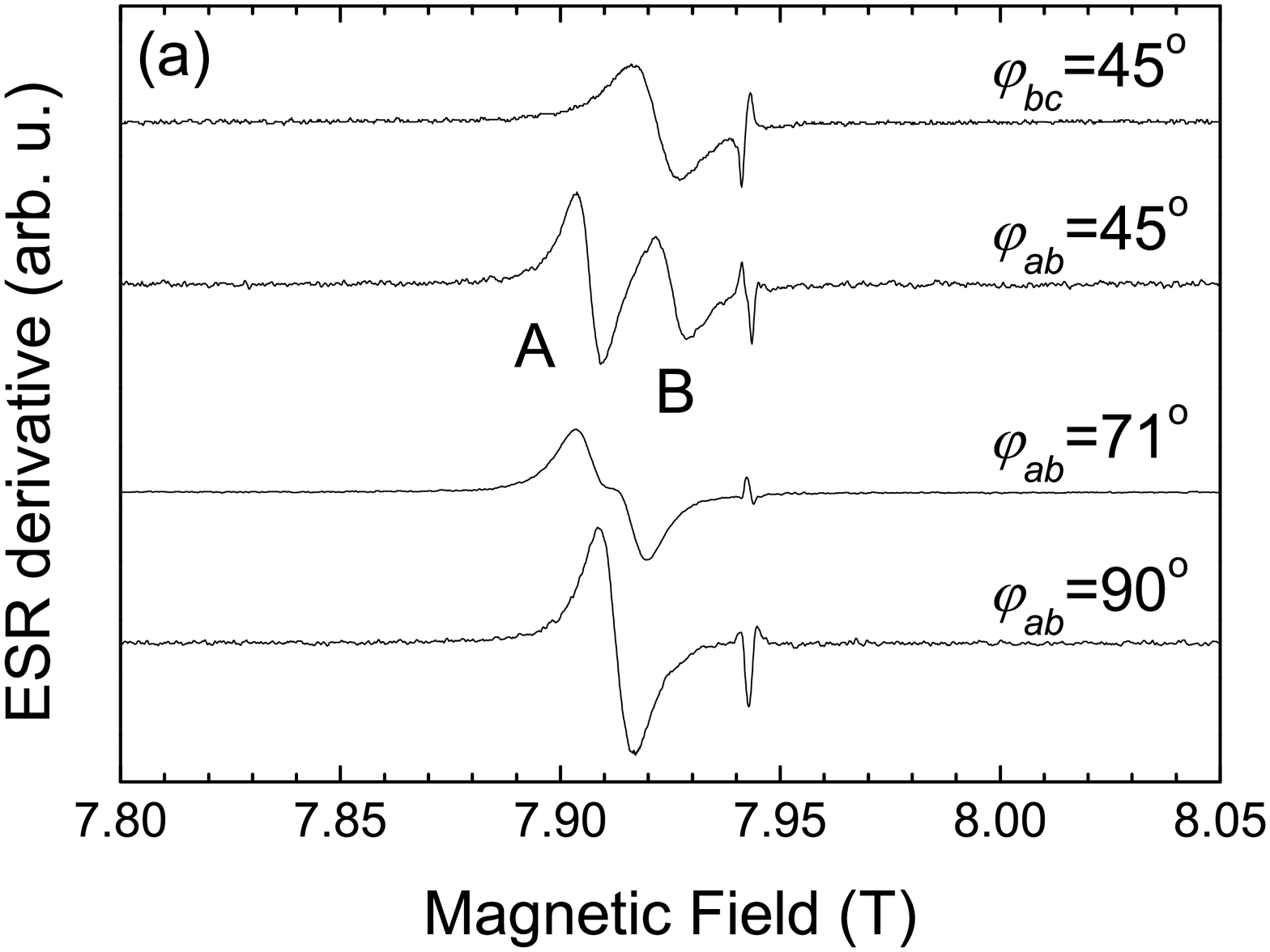}\\
\includegraphics[width=\columnwidth, trim=0 20 0 25,clip]{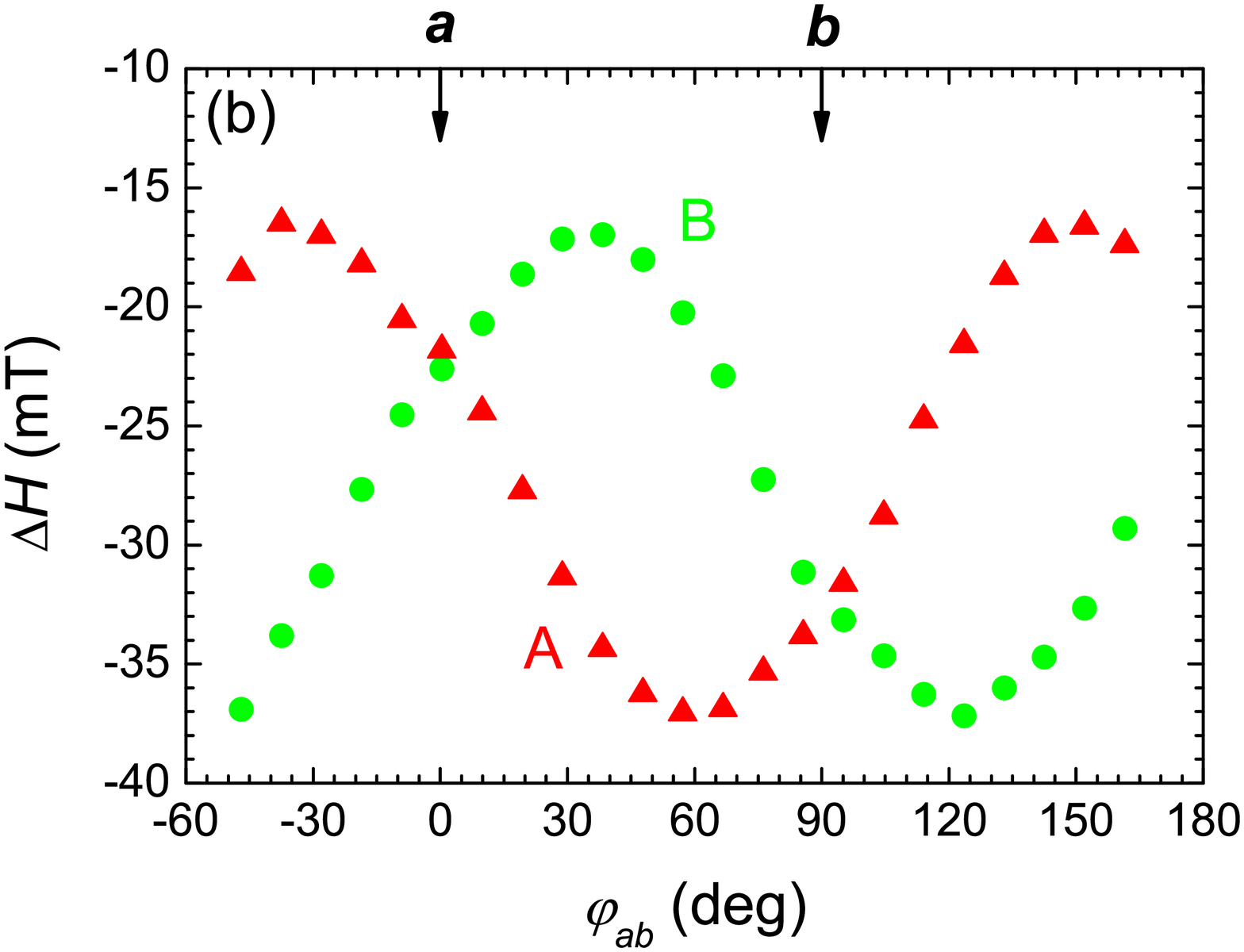}%
\caption{(color online) Conduction electron spin resonance in \kCl\ at
$222.4\rm\,GHz$ and $250\rm\,K$.
(a) Typical derivative CESR spectra. The resolved lines of A and B
layers at fields in general directions ($\varphi_{ab}=45^\circ$ and
$\varphi_{ab}=71^\circ$) prove two-dimensionality of spin diffusion.
When the A and B layers are magnetically equivalent
($\varphi_{ab}=90^\circ$ and $\varphi_{bc}=45^\circ$) a single line
appears. It follows from the $4\rm\,mT$ line splitting at
$\varphi_{ab}=71^\circ$ that inter-layer spin diffusion is blocked for
at least $1.4\rm\,ns$. The KC$_{60}$ reference at $H_0=7.94\rm\,T$ has a
$g$ factor of $g_0=2.0006$.
(b) Angular dependence of CESR shift $\Delta H=H_0(g_0-g)/g$ in the
$(\bm{a},\bm{b})$ plane. The principal axes of the $g$-factor tensors in
the A and B layers are tilted from the orthorhombic $\bm{a}$ and
$\bm{b}$ axes. Spectra and $g$-factor anisotropies of Cl and Br
compounds are similar.\label{fig:cesr}}
\end{figure}

Surprisingly, CESR lines of the inequivalent A and B layers in \kX\ 
are clearly resolved. In the $222.4\rm\,GHz$ CESR spectra
[Fig.~\ref{fig:cesr}(a)] at magnetic fields in general directions, two
equal intensity lines appear with differently oriented $g$-factor
tensors, $g_{\rm A}$ and $g_{\rm B}$. The assignment of these lines to
layers A and B respectively is unambiguous. There is a single line at
magnetic fields in the $(\bm{a},\bm{c})$ and $(\bm{b},\bm{c})$ planes,
where A and B layers are magnetically equivalent, but the line is split
in all other directions. A common principal axis of $g_{\rm A}$ and
$g_{\rm B}$ coincides with the $\bm{c}$ crystallographic axis while the
other two are rotated from $\bm{a}$ and $\bm{b}$ by $-30^\circ$ and
$+30^\circ$ for $g_{\rm A}$ and $g_{\rm B}$, respectively
[Fig.~\ref{fig:cesr}(b)]. Overlap between A1 and A2 dimers within
metallic layers is large, typically $t_\parallel=0.1\rm\,eV$, more than
sufficient to merge their ESR into a single line. The $g$-factor
anisotropy of a single \kCl\ layer resembles that of
$\kappa$-(ET)$_2$Cu$_2$(CN)$_3$ \cite{Komatsu1996}, which has a
monoclinic crystal structure with one type of layer. Results were
reproduced for several samples and were similar for $X$=Cl and Br.
Twinning was excluded by X-ray diffraction. The splitting is temperature
independent and is proportional to the frequency for 111.2 and
$222.4\rm\,GHz$. We find no CESR splitting at $9\rm\,GHz$.

The $4\rm\,mT$ minimum observable splitting of CESR lines, corresponding
to $\Delta\omega=7\cdot10^8\rm\,s^{-1}$, implies that inter-layer
hopping is extremely rare: diffusion is two dimensional and spins are
confined to a single molecular layer for $\tau_{\rm spin}\ge
1/\Delta\omega= 1.4\rm\,ns$. (The $T_1 = 1\rm\,ns$ spin lifetime, as
determined from the CESR line width, gives a similar lower limit for
inter-layer hopping.) Spins diffuse to a distance $\delta_{\rm
s}=\frac{1}{2}v_{\rm F}(\tau_\parallel\tau_{\rm
spin})^\frac{1}{2}\ge0.2\rm\,\mu m$ without hopping to the next
molecular layer. Here $v_{\rm F}=10^5\rm\,m/s$ is the Fermi velocity
\cite{Kovalev2003}, and $\tau_\parallel\ge10^{-14}\rm\,s$ since the mean
free path, $l=v_{\rm F}\tau_\parallel$, exceeds the average dimer--dimer
distance of $10^{-9}\rm\,m$.

The transverse charge hopping time is
$\nu_\perp^{-1}=\hbar^2/(2t_\perp^2\tau_\parallel)$ in the incoherent
hopping limit \cite{Kumar1992}, thus the charge confinement is a
consequence of the small inter-layer overlap energy, $t_\perp$, and the
very short intra-layer scattering time,
$\tau_\parallel\ll\hbar/t_\perp$. In a simple metallic picture
transverse spin and charge hopping times are equal, and $\tau_{\rm
spin}\ge1.4\rm\,ns$ implies $\hbar/t_\perp\ge5\cdot10^{-12}\rm\,s$ for
\kCl. This is the same order of magnitude as
$\hbar/t_\perp\ge16\cdot10^{-12}\rm\,s$ measured by magnetoresistance at
low temperatures in a similar compound, $\kappa$-(ET)$_2$Cu(NCS)$_2$
\cite{Singleton2002}. We expect that a moderate increase of $t_\perp$ or
of $\tau_\parallel$ (e.g.\ at lower temperature or under pressure)
breaks the confinement of spin diffusion to a single molecular layer. An
unexplained line broadening below $50\rm\,K$ prevents testing this in
\kBr. The low inter-layer hopping rate suggests a much larger
conductivity anisotropy $\sigma_\parallel/\sigma_\perp=
(t_\parallel/t_\perp)^2(l/b)^2= 10^5$--$10^6$ than is measured by dc
techniques \cite{Buravov1992}. Leakage through defects of the insulating
polymer may explain this discrepancy \cite{McGuire2001}, although it
would result in CESR intensity between the A and B lines, which we do
not observe.


In the following we show that in the magnetically ordered state of
\kCl\ the magnetic resonance modes of adjacent layers are independently
excited in magnetic fields in general directions. Unlike the metallic
phase, where perpendicular spin diffusion is not observed, weak
inter-layer interactions inhibit mode crossing in symmetry planes and are
measurable in the magnetically ordered phase. 

The low temperature magnetic structure of \kCl\ is known in detail
from static susceptibility \cite{Welp1992} and NMR
\cite{Smith2004,Smith2003}. A magnetic moment of about
$0.5\muB$ resides on each dimer. Layers A and B are slightly canted
two-sublattice antiferromagnets with large intra- and small inter-layer
exchange energies, $\lambda$ and $\lambda_{\rm AB}$, respectively. The
Dzyaloshinskii--Moriya vectors, $\bm{D}_\xi$, characterizing the weak
ferromagnetism are differently oriented in the $\xi=\rm A$ and B layers:
they lie in the $(\bm{a},\bm{b})$ plane at $\varphi_{\rm A}=46.5^\circ$
and $\varphi_{\rm B}=133.5^\circ$, respectively, from the $\bm{a}$ axis.

\begin{figure}
\includegraphics[width=\columnwidth, trim=0 30 0 10,clip]{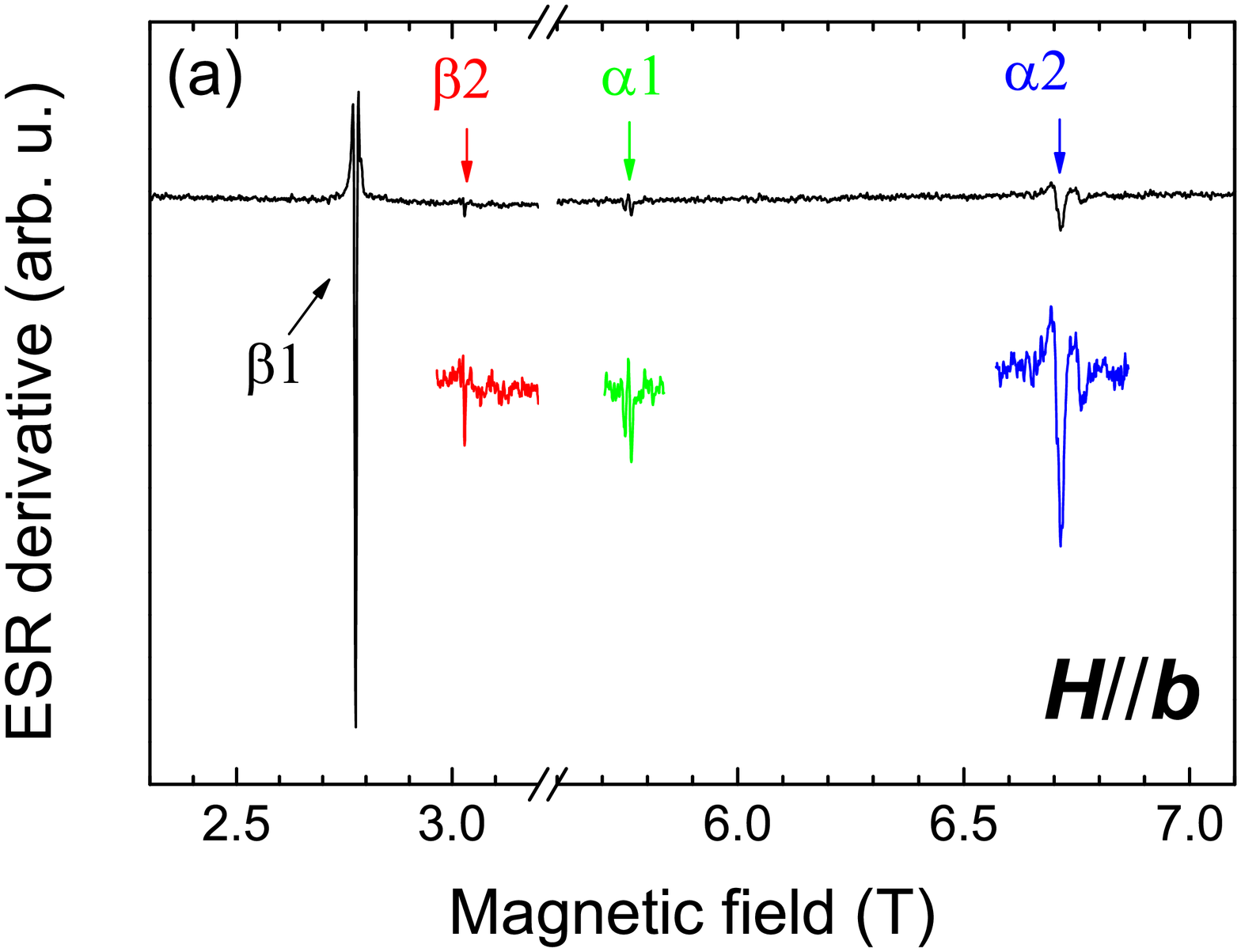}\\
\includegraphics[width=\columnwidth, trim=0 30 0 10,clip]{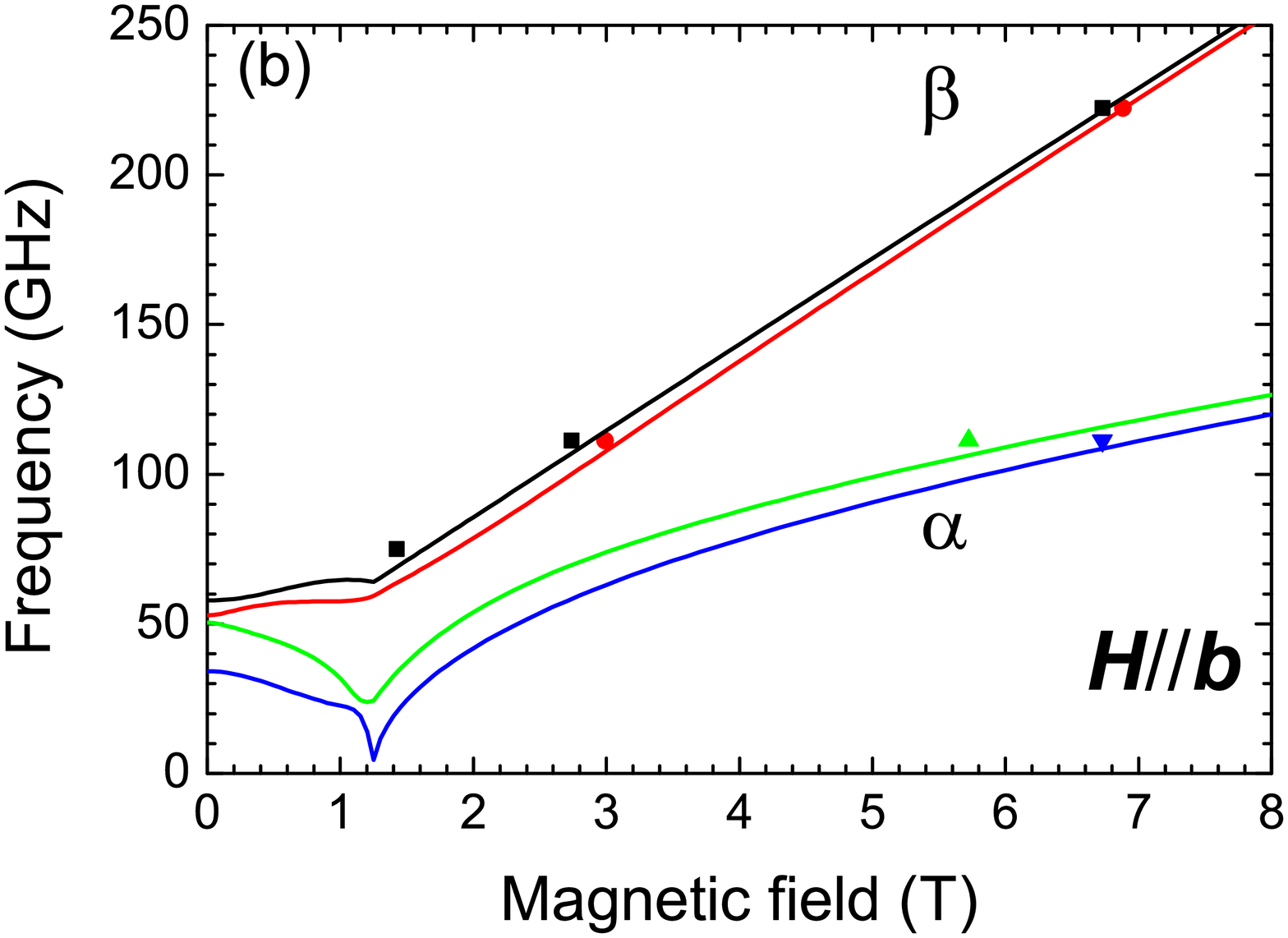}%
\caption{(color online) Antiferromagnetic resonance in \kCl\ at $T =
4\rm\,K$ and $\bm{H}\parallel\bm{b}$. The 4 modes are the $\alpha$
(green, blue) and $\beta$ (red, black) eigenoscillations of weakly
coupled canted antiferromagnetic A and B layers. (a) AFMR spectrum at
$111.2\rm\,GHz$. $5\times$ amplified spectra are shown for higher field
lines. (b) Resonance-field--frequency mode diagram.\label{fig:afmr-b}}
\end{figure}

\begin{figure}
\includegraphics[width=\columnwidth, trim=0 30 0 10,clip]{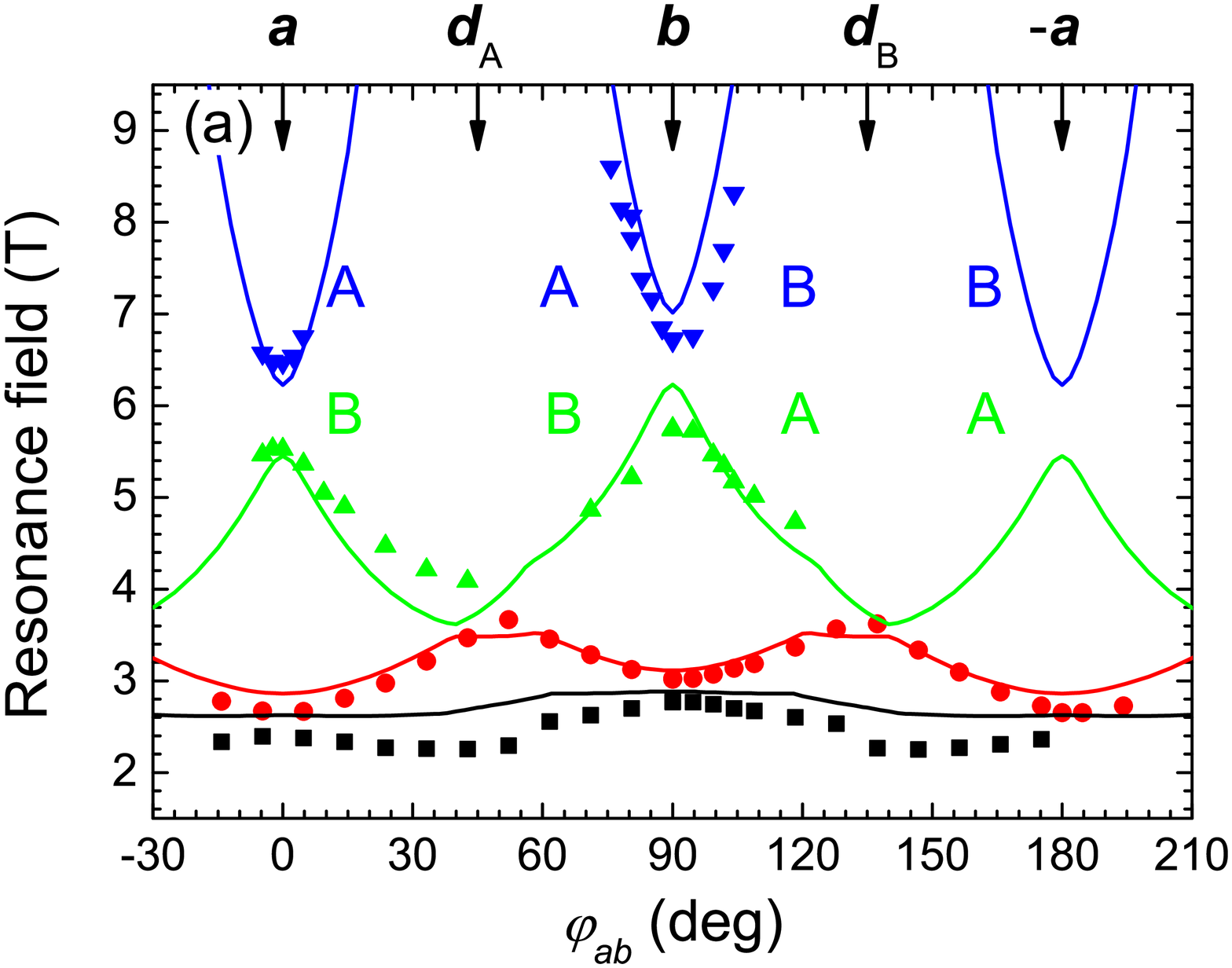}\\
\includegraphics[width=\columnwidth, trim=0 30 0 10,clip]{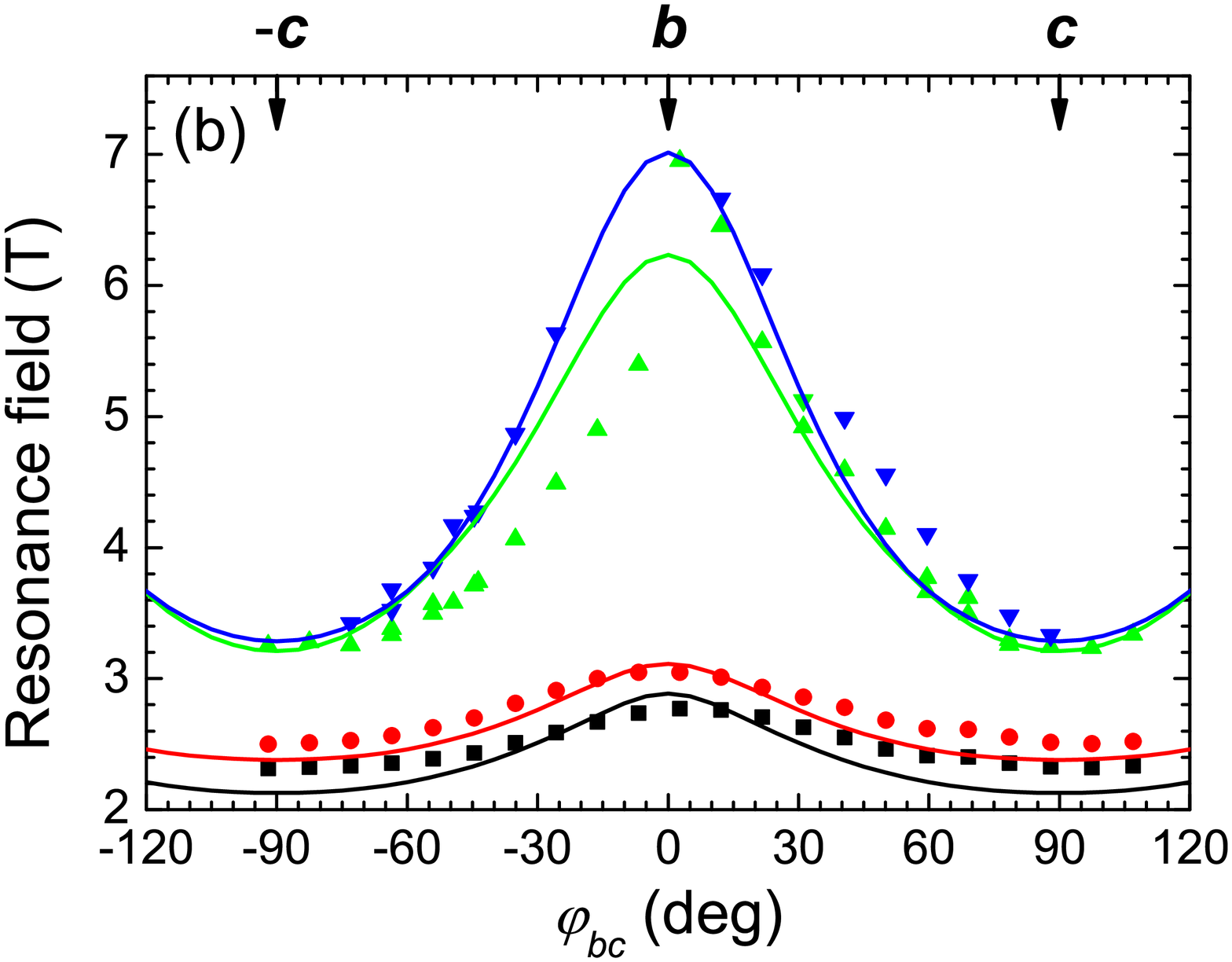}%
\caption{(color online) Observed (dots and squares) and calculated
(lines) AFMR magnetic fields at $111.4\rm\,GHz$. (a) In the
$(\bm{a},\bm{b})$ plane the $\alpha$ modes of A and B layers diverge as
$\bm{H}$ is tilted towards the Dzyaloshinskii--Moriya vectors at
$\varphi_{ab}\approx45^\circ$ and $135^\circ$. They are little affected
by the inter-layer interaction except at mode crossings near $\bm{a}$ and
$\bm{b}$. (b) The degenerate AFMR mode in the $(\bm{b},\bm{c})$ symmetry
plane is split by the inter-layer interaction. The $\pm5^\circ$
uncertainty in sample orientations and approximations in the calculation
described in the text explain differences between experiment and
theory.\label{fig:afmr-ab}}
\end{figure}

The ESR of \kCl\ shifts rapidly with temperature as the
antiferromagnetic order develops below the Mott transition at
$27\rm\,K$. In crystals of the highest quality batch the magnetic
resonance is narrow at low temperatures [Fig.~\ref{fig:afmr-b}(a)]. The
angle dependence of the resonance magnetic field at $4\rm\,K$ was
mapped at frequencies $111.2\rm\,GHz$ (Fig.~\ref{fig:afmr-ab}) and
$222.4\rm\,GHz$ (not shown). The antiferromagnetic resonance (AFMR) data
of Ohta et al.~\cite{Ohta1999} for $\bm{H}\parallel\bm{b}$ agree with
one of our observed modes. The mode diagrams of the frequency versus
resonance field in Fig.~\ref{fig:afmr-b}(b) were calculated numerically
at fixed field directions from the free energy, $F=F_{\rm A}+F_{\rm
B}+F_{\rm AB}$, of weakly interacting adjacent layers and a
four-sublattice dynamics:
\begin{eqnarray}
F_\xi&=&-\bm{H}\cdot\bm{M}_\xi
        -\lambda\bm{M}_{\xi1}\cdot\bm{M}_{\xi2}\nonumber\\
  &&+\bm{D}_\xi\cdot(\bm{M}_{\xi1}\times\bm{M}_{\xi2}) \nonumber\\
  &&+ (K_c/2)[(\bm{M}_{\xi1}\cdot \hat{\bm{c}})^2+
        (\bm{M}_{\xi2}\cdot \hat{\bm{c}})^2 ] \nonumber\\
  &&+ (K_D/2)[(\bm{M}_{\xi1}\cdot \hat{\bm{d}}_\xi)^2+
        (\bm{M}_{\xi2}\cdot \hat{\bm{d}}_\xi)^2 ] \nonumber\\
F_{\rm AB}&=&-\lambda_{\rm AB}(\bm{M}_{\rm A1}\cdot\bm{M}_{\rm B1}+
  \bm{M}_{\rm A2}\cdot\bm{M}_{\rm B2}) \nonumber
\end{eqnarray}
where $\bm{M}_{\xi1}$, $\bm{M}_{\xi2}$ are the sublattice magnetizations
of layer $\xi$ with magnitude $M_0$,
$\bm{M}_\xi=\bm{M}_{\xi1}+\bm{M}_{\xi2}$, and $\hat{\bm{c}}$ is the
$\bm{c}$ axis unit vector. We approximated $\varphi_{\rm A}$ and
$\varphi_{\rm B}$ with $\varphi_{\rm A0}=45^\circ$ and $\varphi_{\rm
B0}=135^\circ$, i.e., we took $\bm{D}_\xi=D\hat{\bm{d}}_\xi$ with
$\hat{\bm{d}}_{\rm A}=[110]/\sqrt{2}$ and $\hat{\bm{d}}_{\rm
B}=[\bar{1}10]/\sqrt{2}$. $K_c$ and $K_D$ denote single-ion anisotropies
with principal axes along $\bm{c}$ and (for simplicity)
$\hat{\bm{d}}_\xi$, respectively. The computer program calculates the
equilibrium directions of $\bm{M}_{\xi1}$ and $\bm{M}_{\xi2}$ first and
then the eigenfrequencies as a function of magnetic field, $\bm{H}$,
oriented in fixed directions. A small damping term towards equilibrium
was also added to obtain finite line widths. The $g$-factor anisotropy
and anisotropic terms in the exchange energies were neglected.

The eigenmodes for a single layer with $\bm{H}\parallel\bm{c}$ are
approximately \cite{Pincus1960}:
\begin{eqnarray}
\omega_\alpha/\gamma&=&\sqrt{DM_0(H+DM_0)+(K_D\lambda M^2)} \nonumber\\
\omega_\beta /\gamma&=&\sqrt{H(H+DM_0)+(K_c\lambda M^2)} \nonumber
\end{eqnarray}
and we denote by $\alpha$ and $\beta$ the continuation of these modes as
the magnetic field angle is varied. The agreement between calculation
and experiment is very good at the various frequencies and for all field
directions (Figs.~\ref{fig:afmr-b} and~\ref{fig:afmr-ab}). The
frequencies depend mainly on $DM_0$ and the products $\lambda_{\rm
AB}\lambda M_0^2$, $K_c\lambda M_0^2$ and $K_c\lambda M_0^2$ but little
on the separate values $\lambda_{\rm AB}$, $K_c$ and $K_D$.
With $\lambda M_0=-450\rm\,T$ (the average value of
Ref.~\cite{Smith2004}), the best fit parameters are $DM_0=3.7\rm\,T$,
$K_DM_0=-11\rm\,mT$, $K_cM_0=2.5\rm\,mT$, and $\lambda_{\rm
AB}M_0=1.15\rm\,mT$. The magnitude of $D$ determined from AFMR and NMR
\cite{Smith2003} agree satisfactorily.

The qualitative features predicted for a pair of weakly coupled
antiferromagnetic layers are demonstrated by the experiment. At fixed
frequencies the $\alpha$ mode resonance field diverges when $\bm{H}$ is
tilted towards $\bm{D}_\xi$ , while the $\beta$ mode has no divergence.
Close lying pairs of $\alpha$ and $\beta$ modes appear in the
$(\bm{a},\bm{c})$ and $(\bm{b},\bm{c})$ symmetry planes. In these planes
the interaction between the A and B layers prevents mode crossing; the
splitting depends largely on $\lambda_{\rm AB}\lambda M_0^2$, but little
on the excitation frequency or angle. The extreme smallness of the
inter-planar interaction, $\lambda_{\rm AB}$, is the most important
finding. The magnetic dynamics is two dimensional because $\lambda_{\rm
AB}$ is almost 6 orders of magnitude smaller than the in-plane
inter-dimer exchange interaction $\lambda$. We found $\lambda_{\rm AB}$
ferromagnetic but this may not be meaningful as its magnitude is
comparable to the dipolar interaction and we neglected its anisotropy.


In conclusion, at high temperatures spin diffusion in the organic
layered conductors is confined to single molecular layers within the
spin lifetime. This feature is desirable for materials with spintronic
information transfer applications as each molecular layer may serve as
an independent channel. Perpendicular transport is strongly incoherent
and both superconducting \kBr\ and antiferromagnetic \kCl\ are
two-dimensional (and not simply anisotropic) metals at high
temperatures. In the ordered magnetic phase the dynamics follows
magnetic eigenoscillations of nearly independent single layers;
inter-layer exchange interactions are comparable or smaller than magnetic
dipolar energies. The observed weakly coupled AFMR modes in \kCl\ 
confirms the microscopic model of Smith et
al.~\cite{Smith2004,Smith2003} and resolves a long standing enigma
\cite{Ohta1999,Ito2000}. It remains to be seen how the inter-layer
coupling changes under pressure.

\begin{acknowledgments}
We are thankful to Dr.~M\'aty\'as Czugler and Veronika Kudar (Chemical
Research Inst.\ Budapest) for X-ray diffraction studies and to
N.~D.~Kushch (Inst.\ of Problems of Chemical Physics, Chernogolovka,
Russia) for instructions on crystal growth. This work was supported by
the Hungarian National Research Fund OTKA NK60984, PF63954, K68807.
T.~F.\ acknowledges the J\'anos Bolyai Research Scholarship of the
Hungarian Academy of Sciences.
\end{acknowledgments}


\end{document}